\documentclass{ametsocV5}

\usepackage{graphicx}
\usepackage{natbib}
\usepackage{xfrac}
\usepackage{comment}
\usepackage{siunitx}
\usepackage{amsmath}
\usepackage{subcaption}
\usepackage{xcolor}
\usepackage{siunitx}
\usepackage{makecell}
\usepackage{booktabs}

\newcommand{\Volume}{{\ooalign{\hfil$V$\hfil\cr\kern0.08em--\hfil\cr}}}

\let\realverbatim=\verbatim
\let\realendverbatim=\endverbatim
\renewcommand\verbatim{\par\addvspace{6pt plus 2pt minus 1pt}\realverbatim}
\renewcommand\endverbatim{\realendverbatim\addvspace{6pt plus 2pt minus 1pt}}
\makeatletter
\newcommand\verbsize{\@setfontsize\verbsize{10}\@xiipt}
\renewcommand\verbatim@font{\verbsize\normalfont\ttfamily}
\makeatother

\nolinenumbers
\title{Impact of stratification mechanisms on turbulent characteristics of stable open-channel flows\footnote{This work has not yet been peer-reviewed and is provided by the contributing author(s) as a means to ensure timely dissemination of scholarly and technical work on a noncommercial basis. Copyright and all rights therein are maintained by the author(s) or by other copyright owners. It is understood that all persons copying this information will adhere to the terms and constraints invoked by each author's copyright. This work may not be reposted without explicit permission of the copyright owner.} \footnote{This work has been submitted to Journal of the Atmospheric Sciences. Copyright in this work may be transferred without further notice.}}
\authors{Cheng-Nian Xiao and Inanc Senocak\correspondingauthor{Inanc Senocak, senocak@pitt.edu}}
\nolinenumbers
\affiliation{Department of Mechanical Engineering and Materials Science, \\ University of Pittsburgh, 3700 O'Hara St, Pittsburgh, PA 15261, USA }

\nolinenumbers
\abstract{Flow over a surface can be stratified by imposing a fixed mean vertical temperature (density) gradient profile throughout or via cooling at the surface. These distinct mechanisms can act simultaneously to establish a stable stratification in a flow. Here, we perform a series of direct numerical simulations of open-channel flows to study adaptation of a neutrally stratified turbulent flow under the combined or independent action of the aforementioned mechanisms. We force the fully developed flow with a constant mass flow rate. This flow forcing technique enables us to keep the bulk Reynolds number constant throughout our investigation and avoid complications arising from the acceleration of the bulk flow when a constant pressure gradient approach were to be adopted to force the flow instead. When both stratification mechanisms are active, the dimensionless stratification perturbation number emerges as an external flow control parameter, in addition to the Reynolds, Froude, and Prandtl numbers. We demonstrate that significant deviations from the Monin-Obukhov similarity formulation are possible when both types of stratification mechanisms are active within an otherwise weakly stable flow, even when the flux Richardson number is well below 0.2. An extended version of the similarity theory due to \citeauthor{zilitinkevich2000} shows promise in predicting the dimensionless shear for cases where both types of stratification mechanisms are active, but the extended theory is less accurate for gradients of scalar. The degree of deviation from neutral dimensionless shear as a function of the vertical coordinate emerges as a qualitative measure of the strength of stable stratification for all the cases investigated in this study.}

\begin{document}

\maketitle

\section{Introduction}

Stable stratifications of the atmospheric boundary layer (ABL) has been the subject of numerous theoretical, computational and experimental studies. They are commonly present in nocturnal boundary layer flows as well as in cold regions such as in the Arctic and Antarctica. It is widely acknowledged that representation of stable conditions in computer models of weather prediction is far from satisfactory \citep{fernando2010_bams, sandu2013, Holtslag2013_bams}.

Stable background stratification of the upper atmosphere, which will henceforth be referred to as \textit{ambient stratification},  
occurs during evening transition by radiative clear-air cooling due to emission by water-vapor, carbon dioxide, ozone and aerosols \citep{andre1982,ha2003}; it  is an important component of the stable atmospheric boundary layer (SABL) in regions of small latitudes close to the poles where  stable atmospheric conditions can be long-lasting as described in various studies
\citep{kitaigorodskii1988,kitaigorodskii1988joffre, zilitinkevich2000,zilitinkevich2002,Zilitinkevich2005}. 
\citet{zilitinkevich2000} have argued that there is a decoupling between surface flow dynamics and the free atmosphere due to the lack of a neutral residual layer in cold climates where the stable atmosphere is long-lived in contrast to warmer regions.
As described in \citet{brunt2011} and \citet{andre1982}, the entrainment of
cold air  over a sloped surface with or without cooling is another configuration in which independent stable ambient stratification can arise.
\citet{kitaigorodskii1988} as well as \citet{kitaigorodskii1988joffre} explored the effect of ambient stratification on  the ABL height to construct an extended similarity theory, which was, however, not validated by experimental or numerical data. \citet{Mironov2010} reviewed different propositions for the ABL height and showed that SABL depth becomes independent of the Coriolis parameter for SBLs dominated by independent ambient stratification as opposed to SBLs dominated by surface buoyancy flux.

Despite the significance of stably stratified conditions, independent stratification of the bulk or free flow has not received much attention in numerical studies published to date. A large majority of numerical simulations of stably stratified flows over flat terrain assumes that the ambient stratification is solely achieved by cooling of the surface, which is also the core assumption in the Monin-Obukhov similarity theory (MOST) as well. This approach links the surface boundary condition directly to the resulting stratification of the flow, as exemplified in the works by \cite{ flores2011, deusebio2014,Gohari2018}. Wind-tunnel experiments of stable boundary layers have also mostly focused on flows stratified solely via cooling of the surface or heating of the top wall, such as described in \cite{ohya1997,ohya2001, williams2017}.
Notable exceptions to purely surface flux-based stratification are the numerical studies of stably stratified Ekman layers conducted by  \citet{taylor2008,taylorsarkar2008}  which included independent ambient stratification of the atmosphere in addition to the surface cooling.  The scaling of wind profiles in Ekman layers stabilized purely by background stratification of the atmosphere has been studied via large-eddy simulations in \citet{kelly2019}.
\citet{fedorovich2017b} prescribed the ambient stratification in the context of gentle slopes and their role in the formation of nocturnal jets that are unique to the U.S. Great Plains. Spatially uniform, time varying buoyancy flux or buoyancy was imposed at the surface. Their study showed that the advection of the ambient stratification due to the sloping terrain can lead to regeneration of turbulence and change in the nocturnal low-level jets.

While it is commonly accepted that fluxes and  gradients in weakly stable flows  are governed by MOST \citep{grachev2015}, numerical studies supporting this hypothesis have hitherto focused on weakly stable flows stratified by surface cooling only \citep{shah2014,deusebio2014}.
The current consensus in the SABL research community is that MOST fails for strongly stratified flows where
turbulence is significantly inhibited and wave motions become a dominant feature of the flow \citep{ohya1997,mahrt2014, williams2017}.
To the best of our knowledge, the
impact of multiple stratification mechanisms on the scaling of flow fields and applicability of MOST as theorized in \citet{kitaigorodskii1988,zilitinkevich2000}, i.e. when stratification is not solely an outcome of surface cooling, has not yet been validated by numerical investigations.

In the present work, we perform direct numerical simulations (DNS) to study the  dynamics of an initially neutral  open-channel flow  at constant bulk Reynolds number subjected abruptly to different strengths of ambient stratification. The open-channel flow configuration has been adopted in the numerical studies by \citet{nieuwstadt2005} and \citet{flores2011} to investigate stably stratified turbulence. The chief distinguishing aspect of our study is the simultaneous existence of two separate stratification mechanisms, one being a prescribed ambient stratification imposed through a constant Brunt-Vais\"al\"a frequency $N_a$ in the governing equations, and the other being a constant cooling buoyancy flux imposed at the bottom surface of the open-channel.
This configuration treats the ambient stratification, which serves to model the free-atmosphere stratification,  independent of the surface buoyancy flux, as suggested in \cite{zilitinkevich2000,kitaigorodskii1988}.
\textcolor{black}{Obviously, in more realistic situations, the independent atmospheric stratification may not be constant throughout; however, a constant ambient stratification as chosen in the present study significantly facilitates our analysis to provide important insights.  }
Experimentally, an independently pre-stratified linear temperature profile was investigated in \citet{ohya2003} without recognizing its implications as a stratification mechanism independent of the surface cooling applied in their experiments. Note that in prior works, \cite{ohya1997} and \cite{ohya2001} had adopted a uniform temperature profile that was stratified through surface cooling only.
The significance of our approach is that due to the presence of a prescribed ambient stratification, which is in addition to and independent of the surface cooling, dynamics of flow stabilization is no longer controlled by a surface flux characterized by the Obukhov length $L$ alone. Our flow configuration is hence more general than the configurations that have been investigated in other numerical studies  \citep{legaspi_waite_2020,armenio2002channel,flores2011,taylor2005}, in which one of the two stratification mechanisms have been adopted without realizing that the type of stratification mechanisms do affect turbulent characteristics. 

In what follows, we investigate how turbulence statistics are modified by the presence of separate stratification mechanisms through DNS at constant bulk Reynolds number via  constant mass flow rate forcing. 
We will  study how MOST, which is generally accepted to hold for weakly stable flows \citep{nieuwstadt1984, grachev2015}, performs when different stratification mechanisms are at play. 
We also consider an extension to MOST as proposed in \citet{zilitinkevich2000}, which we refer to it as the extended similarity theory, and assess it based on our DNS results.

\section{Technical Formulation}

Let us consider an open-channel flow configuration with heat transfer and stratification, where
$x$ be the streamwise, $z$ be the cross-stream (transverse), and $y$ be the vertical direction in which the gravity acts upon, such that $(x_i)=[x, y, z]$ is the position vector in Cartesian representation. The $x-z$ plane at $y=0$ is the surface wall. Similarly, $u,v,w$ are identified with the streamwise, vertical, and cross-stream (transverse) velocity components, such that $u_i=[u, v, w]$ is the  Cartesian representation of the velocity vector. The normalized gravity vector is given by  $g_i=(g_1,g_2,g_3)=[0, 1, 0]$. Note that the gravity vector is aligned with the $y$ coordinate and not with the $z$ coordinate direction.

Let us also assume that the open-channel flow can exchange thermal energy across its boundaries as sensible heat only. The potential temperature and buoyancy fields are denoted by $\theta$, $b$, respectively. The buoyancy is related to the potential temperature as follows 
\begin{align}
    b=\frac{g}{\Theta_r}(\Theta - \Theta_e), \label{eq:buoyancy}
\end{align}
where $\Theta_{r}$ is a reference potential temperature and $\Theta_e$ is the environmental or the background potential temperature that is assumed to be steady in time and vary with height only. The variation of the environmental potential temperature $\Theta_e$ creates an ambient stratification characterized by the  Brunt-V\"ais\"al\"a or buoyancy frequency as follows:
\begin{align}
N_a=\sqrt{\frac{g}{\Theta_r}\frac{\partial \Theta_e}{\partial y}}.\label{eq:bruntvaisala}
\end{align}
In our configuration, a  constant negative buoyancy flux with magnitude $B_s$ is imposed on the surface as well
\begin{align}
B_s=\beta G_w,
\end{align}
where $\beta$ is the constant thermal diffusivity of the fluid and $G_w=\frac{\partial b}{\partial y} \Big|_{\substack{y=0}} >0 $ is the gradient of buoyancy field at the bottom wall or surface. Thus, there are two independent stratification mechanisms acting simultaneously in our open-channel flow setup, namely a prescribed ambient stratification quantified by $N_a^2$ and a negative buoyancy flux imposed  at the channel's bottom wall.

The canonical nature of our flow configuration enables us to determine the complete set of the dimensionless parameters governing the flow problem at hand. In order to equally account  for the effects of both stratification mechanisms, we introduce a \textit{composite stratification} measure as
\begin{align}
     N_c^2 \equiv N_a^2 +G_w.\label{eq:Nc}
 \end{align}

We use the following scaling parameters to nondimensionalize the governing equations: 
 \begin{align} x_i \sim H, \quad u_i \sim U, \quad P_o \sim \rho U^2, \quad t_o \sim H/U, \quad b_0 \sim N_c^2 H \label{eq:scales},
\end{align}
where $H$ is the height of the open channel, $U$ is the bulk velocity in the open channel, $P_o$ is the reference pressure, $\rho$ is the density of the working fluid, $t_o$ is the time scale, and $b_o$ is the buoyancy scale. We, then, write the dimensionless form of the conservation of momentum and the buoyancy balance equations with a Boussinesq approximation as follows:
\begin{align}
\frac{\partial u_i}{\partial t}+\frac{\partial u_i u_j}{\partial x_j}
=&~ -\frac{\partial p}{\partial x_i} + 
\frac{1}{Fr_c^2}b \delta_{i2}+ \frac{\partial}{\partial x_j} \frac{1}{Re_b}\left( \frac{\partial u_i}{\partial x_j}\right), \label{eqnslopemom} \\
\frac{\partial b}{\partial t} +  \frac{\partial u_j b}{\partial x_j}  = &~   \frac{1}{Re_b\ Pr}\frac{\partial}{\partial x_j}\left( \frac{\partial b}{\partial x_j} \right)  -  \frac{1}{1+\Pi_s}\delta_{j2} u_j \label{eqnslopebuoy},
\end{align}
where $\delta_{ij}$ is the Kronecker delta function and all quantities in the above equations are dimensionless variables. Dimensionless numbers appearing in the above equations are defined as follows:
\begin{align}\label{eq:pi_set}
     Pr \equiv \frac{\nu}{\beta} \quad Re_b \equiv \frac{UH}{\nu}, \quad Fr_c \equiv \frac{U}{H N_c}, \quad \Pi_s \equiv \frac{|G_w|}{N_a^2},
\end{align}
where $Pr$ is the Prandtl number, $\nu$ is the kinematic viscosity, $\beta$ is the thermal diffusivity, $Re_b$ is the Reynolds number based on the bulk velocity, $Fr_c$ is the \textit{composite} Froude number based on the composite stratification $N_c$ as introduced in Eq. (\ref{eq:Nc}), and $\Pi_s$ is the stratification perturbation parameter \citep{xiao2019,senocak2020}. At the surface, the normalized buoyancy gradient is related to the stratification perturbation number as follows: 
\begin{equation}
G_w=\frac{\partial b}{\partial y}\biggr|_{y=0} = \frac{\Pi_s}{1+\Pi_s}.  \label{eqnsurface}
\end{equation}

From Eq. \eqref{eqnslopebuoy}, it can be seen that  the stratification perturbation parameter $\Pi_s$ would disappear  after ensemble-averaging since the mean vertical velocity in a channel flow is zero; however,  the mean buoyancy gradient does contribute to the production of turbulent buoyancy fluxes and variances \citep{shah2014}. This suggests that $\Pi_s$ would have no influence in the laminar case, but it does impact the turbulent stratified flow.

In addition to Eqs. (\ref{eqnslopemom}) and (\ref{eqnslopebuoy}), the conservation of mass principle is enforced  by a divergence-free velocity field
\begin{equation}
    \frac{\partial u_i}{\partial x_i}=0.\label{eqnslopecont}
\end{equation}

The \textit{composite} Froude number can be further decomposed into contributions from the bulk flow ($Fr_b$) and surface boundary conditions ($Fr_w$) as follows: 
\begin{align}
\frac{1}{Fr_c} \equiv \sqrt{\left(\frac{1}{Fr_b^2}+ \frac{1}{Fr_w^2}\right)},\label{eq:compFroude}
\end{align}
where the \textit{bulk} and the \textit{wall} (or \textit{surface}) Froude numbers are defined as $Fr_b = U \ (H N_a)^{-1}$ and  $Fr_w = U \ (H \sqrt{G_w})^{-1}$, respectively. Note that $Fr_b$ and $Fr_w$ do not constitute as two independent dimensionless numbers in the control parameter space when both $N_a$ and $G_w$ are simultaneously active in the flow problem since we use $\Pi_s$ as one of the dimensionless parameters in such cases. Together, $Fr_b$ and $Fr_w$ account for the Froude number effects in the flow through the \textit{composite} Froude number.

\textcolor{black}{
At this point, it is worth mentioning  that other dimensionless parameters  have also been introduced elsewhere for the characterization of stably stratified flows,  such as e.g. the \textit{shear capacity} (SC) defined  by \citet{van2018parameters}, which  can be written as the cubic root of the normalized Obukhov length introduced by \cite{flores2011}. It can be readily shown that these two parameters are algebraically related to the Froude number, Reynolds number and Prandtl number.} 
\textcolor{black}{However, the  presence of an  ambient stratification independent of surface cooling, which is the subject of the current study, necessitates the introduction of a novel dimensionless parameter which characterizes  the strength of this additional stratification. As shown in Eq. \ref{eq:pi_set}, we have defined the stratification perturbation parameter $\Pi_s$ for this purpose as the relative strength of ambient stratification magnitude  with respect to the surface buoyancy gradient.}
After some manipulations, we arrive at the following relations 
\begin{equation}
\Pi_s = \left(\frac{Fr_b}{Fr_w}\right)^2, \quad Fr_c= \frac{Fr_b}{\sqrt{1+\Pi_s}}.
\end{equation}

From the above arguments, it should be clear that the set of dimensionless numbers governing a stratified flow can be defined in several ways. 
In the context of stratified atmospheric flows over flat terrain, we emphasize that ambient stratification have not received much attention as a mechanism of stratification independent of surface cooling. Cooling at the surface imposed either as a negative heat flux or a prescribed temperature has been the primary mechanism of stratification in prior numerical studies of SABL over flat terrain.  When both mechanisms of stratification are taken into consideration, as done in slope flows \citep{fedorovich2009, fedorovich2017b, xiao2020anabatic}, the list of relevant physical parameters for the open-channel flow problem consists of the fluid kinematic viscosity $\nu$ and its thermal diffusivity $\beta$, the characteristic velocity $U_0$, the open channel height $H$, the imposed surface buoyancy flux $B_s$ and the prescribed ambient stratification $N_a$. From the Buckingham  $\pi$-theorem, there are exactly four independent dimensionless parameters characterizing the entire flow configuration. Although, various dimensionless parameters can be manipulated to form different versions, we recommend the following set because it is composed of the more familiar dimensionless numbers while uniquely defining the parameter space of the current problem: $\{Re_b, \quad Fr_b, \quad \Pi_s, \quad Pr\}$. We emphasize that in prior studies of channel-type flows the dimensionless parameter space, excluding the Prandtl number, was composed of only two parameters, i.e. Reynolds number and either Froude or Richardson number \citep{nieuwstadt2005, garcia2011, flores2011}. For completeness of the discussion, we note that if a nonzero buoyancy $b_{\infty}$ or buoyancy flux $B_{\infty}$ were imposed on the top boundary as a special case, then an additional dimensionless parameter characterizing its effect would also appear.

\section{Description of Direct Numerical Simulations} 
We solve the three-dimensional system of incompressible Navier-Stokes equations
using a high order  h/p continuous
Galerkin–Fourier solver available in the open source \textit{Nektar++ 5.0.0} package. \textit{Nektar++} has been extensively validated based on accurate solutions for a series of complex flow problems \citep{ nektar++, moxey2020}.
A second order stiffly stable time integration scheme with constant time step satisfying the CFL $\le$ 0.4 was adopted in the simulations.
For spatial discretization, we use a spectral element mesh which is refined in the vertical direction such that   with p-refinement of each element at polynomial order 10, more than 12 collocation points are contained within the first 10 wall units from the surface.
Along the horizontal directions, we have $\Delta x^+ \approx 12$ and $\Delta x^+ \approx 6$, as recommended in \cite{flores2011}.

All the cases studied hereafter adopt a constant mass flow rate as the flow forcing technique in DNS and initiate from a fully developed unstratified turbulent open channel flow at initial friction Reynolds number \textcolor{black}{ $Re_{\tau 0}=u_{*0}H \nu^{-1} = 360$, where $u_*=\sqrt{\nu\left.\frac{\partial <u>}{\partial y}\right|_{y=0} }$ is the friction velocity, and the subscript $0$ has been used to refer to the value of the respective quantity at initial time $t=0$.} This condition is equivalent to a bulk Reynolds number of $Re_b=12,200$. The imposition of constant flow rate, rather than a constant pressure gradient as done in other numerical studies, greatly reduces flow acceleration within channel core during the stabilization process. We provide further details about constant flow rate forcing technique in \S \ref{sec:bulkdp} in the Appendix and compare it against the constraint mean pressure gradient forcing technique. 
The Prandtl number in all cases is fixed to 0.71. Following the setup described in \cite{flores2011}, we set the horizontal domain size in all cases to $8\pi H \times 4\pi H$, where $H$ is the open channel's height.  As pointed out in \citet{garcia2011}, the domain size for stratified channel flows needs to  be sufficiently large in order to prevent artificial oscillations of second-order statistics at higher stability. It has also been hypothesized in \citet{garcia2011,coleman1990} that a simulation domain which is insufficient to resolve the larger structures of stable flows will lead to premature relaminarization.
As a result, due to the necessity of employing larger domains, the computational cost for stable flows even at small Reynolds numbers can quickly become substantial, which was the primary reason for our choice of the moderate value $Re_{\tau}=360$. 

In departure from other fundamental studies on stratified flows over flat terrain \citep{flores2011, taylor2005, nieuwstadt2005}, we prescribe a constant ambient stratification ($N_{a}^{2}$) on the flow field in addition to a negative buoyancy (heat) flux ($B_s$) imposed at the surface. In other words, we prescribe a stratification perturbation number $\Pi_s$ to each simulation as an external control parameter. We investigate three cases with different combinations of independent ambient stratification and surface buoyancy flux to elucidate how different stratification mechanisms modify the flow stabilization process as well as the stationary stratified flow profiles and investigate their impact on MOST. 

We configure the  simulation cases as follows: Our first configuration \textit{Case} {I} is purely stabilized  by a constant homogeneous cooling flux $B_s$ at the bottom surface.  The next configuration \textit{Case} {II} is obtained by keeping the same surface flux as \textit{Case I}, \textcolor{black}{but in addition we impose a nonzero $N_{a,II}^2$ as the  prescribed ambient stratification in the open channel.} 

To investigate the sole influence of an independent, constant ambient stratification on turbulent flows, we setup \textit{Case III} with the same  ambient stratification $N_{a,III}^2=N_{a,II}^2$ as \textit{Case II} and \textit{Case VI} with a five times stronger value  $N_{a,VI}^2=5 N_{a,II}^2$; both cases have zero surface flux. Finally,
we setup \textit{Case IV} and \textit{Case V} such that they share the same composite Froude number among them. \textit{Case IV} has the same surface buoyancy flux as \textit{Case I,II}, but additionally contains a constant ambient stratification five times higher than the value used in \textit{Case II} (i.e. $N_{a,\text{IV}}^2=5N_{a,\text{II}}^2$). On the other hand, \textit{Case V} is purely stabilized by surface buoyancy flux that is higher than in \textit{Case I}. In all simulations, we adopt no-slip conditions combined with buoyancy flux at the bottom surface ($y=0: u =v =w =0, \frac{\partial b}{\partial y} =B_s\beta^{-1}$).
At the top boundary $y=H$, we apply free-slip conditions for the velocities($v=0, \frac{\partial u}{\partial y}= \frac{\partial w}{\partial y} =0)$. 
\textcolor{black}{We impose adiabatic conditions on the buoyancy field at the top boundary for all cases. It is important to point out that for the open channel top, a Dirichlet type of thermal boundary condition was applied in other numerical studies \citep{flores2011,nieuwstadt2005}. Dirichlet type of boundary condition has a clear impact on the heat transfer characteristics of the flow. In the open-channel flow simulations described in \cite{atoufi2020characteristics}, the effects of different boundary conditions at the channel top have been studied,  and it has been determined that heat entrainment at the top boundary due to the Dirichlet condition can modulate the characteristics of stratified flows such as turbulent kinetic energy, mean shear and stratification. We present the influence of thermal boundary condition at the open-channel top boundary in \S \ref{sec:effectbc} in the Appendix. } 

Dimensionless flow control parameters for all six cases are tabulated in Table \ref{tab:stratifiedparam}. Note that the bulk Reynolds number based on channel height and mean velocity is kept constant at 12,200 through a constant mass flow rate forcing technique. Judging purely from the composite Froude number $Fr_c$ of each configuration shown in Table \ref{tab:stratifiedparam},  \textit{Case III} has the weakest overall stratification, followed by \textit{Case I} which is slightly less stratified than   \textit{Case II}, and cases \textit{ IV,V} are the two equally most stratified configurations. We should mention that none of these cases fall under the ``very stable'' flow regime classification of \cite{mahrt1998} because continuous and fully developed turbulence was sustained in each case. To this end, these flow cases meet the ``weakly stable'' flow regime classification of \citeauthor{Mahrt1989}.

\begin{table}
\centering
\begin{tabular}[t]{@{}lllcccc@{}}
\toprule
   & \thead[bl]{Stratification \\ mechanism} & $Fr_b$ & $Fr_c$& $\Pi_s$ & \thead[bl]{Final \\ $ Re_{\tau}$}  & \thead[bl]{Degree of \\ stability}  \\  \midrule
 Case I \ &$B_s$ only & $\--$ & 1.52 & $\--$  & 335& weak \\ 
 Case II \ & $B_s \ \& \ N_a$ & 6.90 & 1.49 & 20 &  332& weak  \\ 
 Case III \ & $N_a$ only & 6.90 & 6.90 & 0  &357 & very weak \\
 Case IV \ & $B_s \ \& \ N_a$ & 3.02 & 1.36 & 4& 315& moderate  \\
 Case V \ & $B_s$ only & $\--$ & 1.36 & $\--$ &  325& weak  \\
 Case VI \  \ & $N_a$ only & 3.02 & 3.02 & 0&350 & weak  \\
 \bottomrule
\end{tabular}
\caption{Dimensionless parameters for stratified flow cases. All cases have Prandtl number $Pr=0.71$ and an initial friction Reynolds number $Re_{\tau,i}=360$.}\label{tab:stratifiedparam} 
\end{table}

\section{Results and Discussions}
\subsection{Transient flow evolution from neutral turbulence}
\begin{figure}
\centering

\centering
    \includegraphics[width=\textwidth]{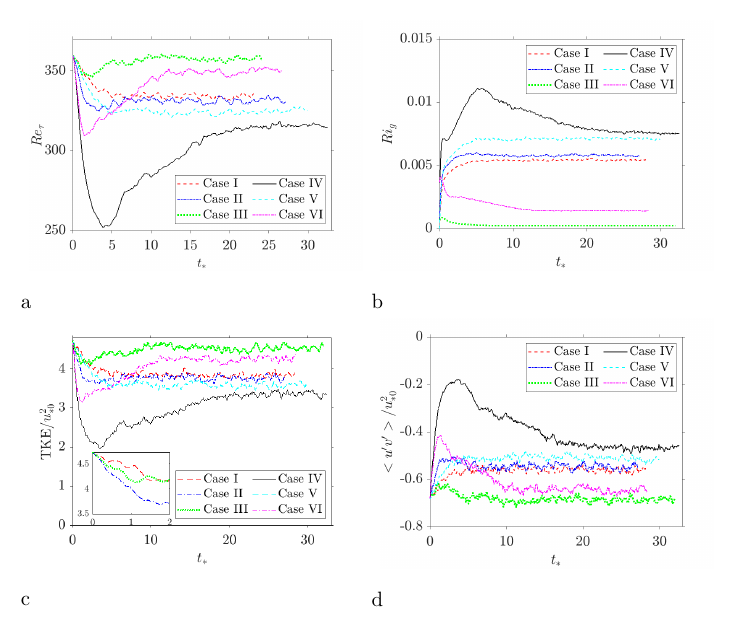}
 
\caption{Evolution of flow quantities from initially neutral flow at $Re_{\tau}=360$ for different combinations of ambient stratification and surface cooling flux, averaged over the horizontal plane: (a) Friction Reynolds number; (b) gradient Richardson number at $y/H=0.05$; (c)  turbulent kinetic energy at $y/H=0.05$ normalized by  square of the initial friction velocity $u_{*0}$ and (d) vertical momentum flux  at $y/H=0.05$ normalized by  $u^2_{*0}$. The height $y/H=0.05$ corresponds to  \textcolor{black}{wall units $y^+=y u_*/\nu= 18$} at initially neutral conditions.    }
\label{fig:evolNhcompare}
\end{figure}

Time evolution of flow statistics for \textit{Cases I-VI} \textcolor{black}{starting from a fully-developed neutral turbulent open channel flow at friction Reynolds number $Re_{\tau} = 360$} is displayed in Figs. \ref{fig:evolNhcompare}a-\ref{fig:evolNhcompare}d. \textcolor{black}{Various plane-averaged turbulent quantities  are shown at the height $y/H=0.05$, which is approximately the location of maximal turbulent kinetic energy, as will be shown later.}
As expected, the effect of stabilizing stratification manifests itself in the decline of friction Reynolds number $Re_{\tau}$,  turbulent kinetic energy (TKE),  Reynolds shear stress  and the growth of gradient Richardson number $Ri_g$ \textcolor{black}{(defined as Eq. \ref{eqnRig})} from their initial neutral values with increasing time. We observe from these plots that for all configurations studied, there is almost an exact inverse relation between  
 flow stability  and composite Froude number $Fr_c$ given in table \ref{tab:stratifiedparam}. The most stable among the six configurations is \textit{Case IV} with $Fr_c=1.36$ where the final $Re_{\tau}$ dropped by over 20\% from the initial $Re_{\tau}=360$ to approximately $Re_{\tau}=284$.
However, even though \textit{Case V} has the same $Fr_c$ as  \textit{Case IV}, Fig. \ref{fig:evolNhcompare}a shows that \textit{Case IV} is significantly more stable than \textit{Case V} whose final $Re_{\tau}$ is larger than 300. This suggests that at the same $Fr_c$, a difference in $\Pi_s$  can significantly impact stability of the turbulent flow, which is in stark contrast to the outcome of the linear stability of laminar flows where stability is solely dependent on $Fr_c$ as described in section \ref{sec:linstable}. Note that bulk Reynolds number $Re_b$ remains constant because of flow forcing via the constant flow rate technique. To summarize, at the same $Fr_c$, the larger the inverse $\Pi_s$, the more stable the flow is. Note that when either the surface flux $B_s$ or the imposed ambient stratification $N_a$ is zero, $\Pi_s$ does not appear in the set of dimensionless numbers.

The weakest stably stratified flow is \textit{Case III} with $Fr_c=6.9$, which attains quasi-equilibrium values with turbulent flow quantities that are nearly indistinguishable from those of the initially unstratified neutral flow. \textcolor{black}{However, we would like to point out that as shown in \cite{atoufi2020characteristics}, the near-wall turbulence quantities of stratified flows at the quasi-stationary state is always very close to the neutral values regardless of the strength of stratification. }
It is worthy to note from  Fig. \ref{fig:evolNhcompare}c that \textit{Case III} exhibits an initial decay of  turbulent kinetic energy for $t_*<1$, only to recover later to almost its starting values. Hence, we conjecture that this abrupt initial turbulence decay is a signature of ambient stratification imposed on the flow, which is further supported by comparing the transient flow data of  \textit{Case I} and \textit{Case II}. The only difference among those two configurations is the presence of a relatively weak prescribed ambient stratification corresponding to  $\Pi_s=20$ in \textit{Case II}, which only has 5\% the magnitude of the same surface buoyancy flux present in both \textit{Case I} and \textit{Case II}. However, the same rapid initial turbulence decay can be observed  in Fig. \ref{fig:evolNhcompare}c for \textit{Case II}, but is absent for  \textit{Case I}. Both  \textit{Case I} and  \textit{Case II}, however, eventually reach very similar stationary values for turbulent flow quantities, albeit \textit{Case II} does obtain slightly lower values due to the additional presence of a prescribed ambient stratification with a strength of $N_a^2$. Thus, we may conclude that  the presence of even a weak prescribed ambient stratification $N_a^2$ relative to the surface buoyancy flux $B_s$ may substantially accelerate the initial transient turbulence decay, though eventually the  stratified turbulence in quasi-equilibrium will not be altered by much if $\Pi_s\gg 1$, which implies that surface buoyancy flux is much stronger than the prescribed ambient stratification. In contrast to the  other two cases where $\Pi_s\gg 1$, i.e. where surface flux is the dominant stabilization mechanism,   \textit{Case IV} and \textit{Case VI} display a recovery of turbulent shear stress levels starting from $t_* =1.5$ and $t_*=4$, respectively, after a  period of initial decay.
This seems to suggest turbulence enhancement (not to be confused with turbulence collapse and rebirth) at constant flow rate is another hallmark of sufficiently strong ambient stratification where $\Pi_s$ approaches the order of one.

\subsection{Analysis of mean velocity profiles}
 
\begin{figure}
\centering

\centering
    \includegraphics[width=\textwidth]{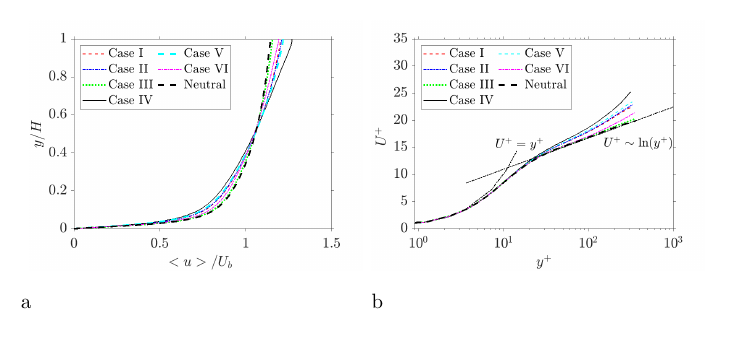}

\caption{Plane-averaged vertical flow profiles at quasi-stationary state for different combination of stratifications, evolving from initially neutral $Re_{\tau}=360$ flow: (a) Streamwise velocity normalized by the bulk velocity; (b) dimensionless mean velocity profile  in wall units \textcolor{black}{ ($y^+=y u_*/\nu, U^+=U/u_*$) } Black dashed dotted line represent the law-of-the wall.}
\label{fig:statmeanub}
\end{figure}

 Fig. \ref{fig:statmeanub}a compares the plane-averaged  velocity profiles along the vertical for the cases \textit{I-VI} at $t_*=20$ and shows that the most stable \textit{Case IV} has  significantly lower  mean shear, hence, less \textcolor{black}{vertical momentum transfer} within channel core compared to the other three cases, which can be explained by the presence of a strong ambient stratification corresponding to $Fr_b=3, \Pi_s=4$.
 Comparing \textit{Case I} and \textit{Case II}, we observe that their velocity profiles agree well for almost the entire height of the channel, but starting to diverge from each other from $y/H>0.8$ onwards due to the additional presence of ambient stratification in \textit{Case II} which further inhibits \textcolor{black}{vertical momentum transfer} compared to \textit{Case I}. 
 The mean velocity profile normalized by the friction velocity is presented in the conventional semi-logarithmic law--of--the--wall (a.k.a. log-law) style in Fig. \ref{fig:statmeanub}b as a function of the wall units. It is clear that all stratified cases  agree with the  linear  profile within the viscous sublayer up until \textcolor{black}{$y^+ := y u_* / \nu \approx 6$}. From $y^+ >20$ onwards it is discernible that \textit{Case IV} with the strongest overall stratification deviates the most from the neutral flow profile due to its stronger mean shear in the channel core caused by suppression of turbulent vertical momentum transfer. To this end, deviations from the theoretical neutral log-law can be used as a qualitative measure of the degree of stable stratification in the flow, i.e.  the larger the deviation from the theoretical neutral log-law, the stronger the degree of stability.
 
\begin{figure}
\centering

    \includegraphics[width=\textwidth]{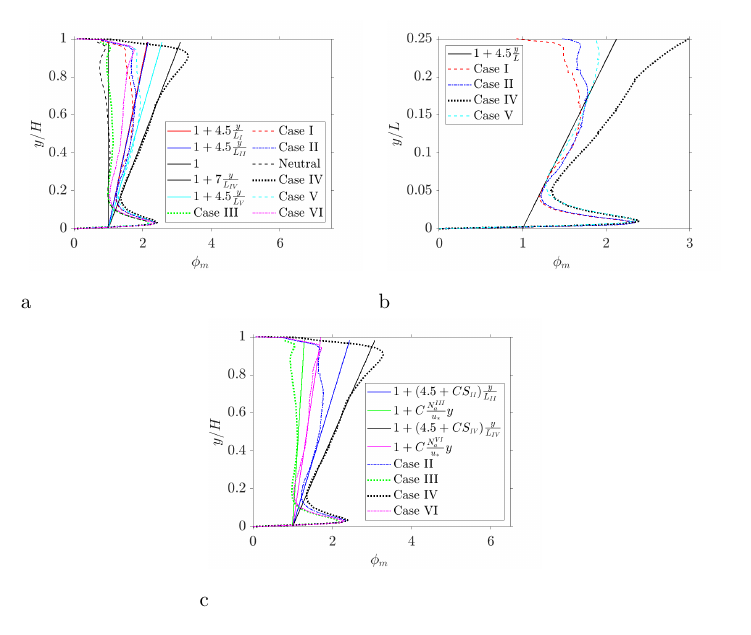}

\caption{Vertical profiles of dimensionless  velocity gradient at quasi-stationary state for different degrees of stratification:
 (a)  as a function of dimensionless height $y/H$ compared to MOST with fitted constant $\alpha_m$; (b)   in terms of the stability parameter $y/L$ compared to MOST with fitted coefficient, excluding the cases without surface buoyancy flux (i.e., \textit{III, VI}) and (c) as a function of dimensionless height $y/H$ for cases with prescribed ambient stratification, compared to extended similarity theory of \cite{zilitinkevich2000} with constant C=0.12.   }
\label{fig:statmeanmo}
\end{figure}

 The dimensionless velocity gradient defined via 
 \begin{equation}
 \phi_m=\frac{\kappa \ y}{u_{*}}\frac{\partial <u>}{\partial y}
 \end{equation}
 is plotted in Fig. \ref{fig:statmeanmo}a for all six stratified cases as well as the initial neutral flow, alongside the
 empirical flux-profile relationships developed from the study of stable atmospheric boundary layers \citep{businger1971,hogstrom1988}: 
 \begin{equation}
 \phi_m=1.0 + \alpha_m\frac{y}{L}\label{eq:phim}
 \end{equation}
 where $\alpha_m$ is an empirical constant, \textcolor{black}{and  $L$ is the Obukhov length defined as $L=-u_*^3/(\kappa B_s)$,  where $\kappa=0.41$ is the von Karman constant.} Different values have been proposed for $\alpha_m$. For example, \cite{hogstrom1988} suggests $\alpha_m=4.8$ as a best fit to empirical data, whereas a value of $\alpha_m=9$ has been obtained from DNS results in \citet{shah2014}. From our DNS data, as presented in Fig. \ref{fig:statmeanmo}a, we observe that the particular value of $\alpha_m$ exhibits a strong dependency on constant ambient stratification imposed on the flow. For cases \textit{ I,V}, which are solely stratified via surface cooling, 
 $\alpha_m=4.5$ is an adequate choice.
 For these two surface buoyancy flux dominated cases with not prescribed ambient stratification (i.e. $N_a=0$), MOST agrees well with the normalized flow gradient profiles at heights satisfying $0.1 ~<~ y/H ~<~ 0.4$. However, for the  stratified configurations \textit{IV,VI} which have significant nonzero ambient stratification imposed throughout the simulation, 
 we observe a very noticeable departure from the  flux-profile relationship which hold for cases \textit{I,V} containing only surface-cooling.
 For \textit{Case IV}, which like \textit{Case VI} possesses strongest magnitude of prescribed ambient stratification(i.e. the lowest bulk Froude number $Fr_b$)   among all configurations,  a  larger value of $\alpha_m=7.0$ was needed to match the DNS data.
 On the other hand, the presence of a much weaker ambient stratification  in case \textit{II} does not appear to cause a discernible deviation from the corresponding  \textit{ Case I}  with the same surface flux magnitude but without prescribed ambient stratification. 
The normalized velocity gradients of the configurations \textit{III,VI} with zero surface cooling flux are expected to follow the log-law just as the initial neutral flow since by definition of the Obukhov length would be infinite. While the log-law approximately holds for \textit{Case III}  due to its weak ambient stratification,  the normalized velocity gradient  of \textit{Case VI} with a stronger ambient stratification appears to increase linearly with height.
These results suggest that deviation  from the linear velocity gradient profile of the pure surface-flux  cases increases with growing strength of the ambient stratification.
 
The stability parameter $y/L$ is generally used as another measure to identify the stability regime of flow at a given height and stratification conditions \citep{wyngaard2010}. 
By definition, the stability parameter for neutrally and stably stratified flows are non-negative, and stability increases with growing $y/L$. 
For $y/L \ll 1$, the flow is very weakly stable and behaves almost like neutral turbulence. On the other end, when $y/L\gg 1$, the flow is very stable such that ``z-less'' stratification is assumed to hold, which means that the flow gradients become independent of height and decoupled from the near-surface turbulence   \citep{nieuwstadt1984}.

In Fig. \ref{fig:statmeanmo}b, where the normalized gradient of the stratified configurations (excluding cases \textit{ III,VI} where the Obukhov length $L$ is ill-defined) is plotted in terms of the stability parameter $y/L$, we observe that all cases start deviating from the linear Monin-Obukhov (M-O) scaling at a weakly stable regime of $y/L \approx 0.1$. This departure from MOST at more stable regimes could be explained by the existence of a free-slip top boundary in the open channel  which might influence  turbulence statistics in higher channel regions where stronger stability is attained.

We note that \cite{zilitinkevich2000} proposed the following extended similarity theory
\begin{equation}
\phi_m = 1.0 + (\alpha_m  + C S) \frac{y}{L} = 1.0 + \alpha_m \frac{y}{L} + C \frac{N_a }{u_*} y  \label{eq:phimextended}
\end{equation}
where $S$ is a dimensionless parameter defined as $S = N_a L u_{*}^{-1}$, and $C$ is an empirical constant taking values between 0.2 and 0.4. \citeauthor{zilitinkevich2000} proposed this extension to MOST to account for the static stability of the free flow.
Based on DNS results of cases \textit{II,III,IV,VI}, all of which have nonzero prescribed ambient stratification, we determined that $C=0.12$ in Eq. (\ref{eq:phimextended}) fits the lower linear portion of all four velocity gradient profiles reasonably well, as shown in Fig. \ref{fig:statmeanmo}c. Thus, our current results suggest that Eq. (\ref{eq:phimextended}) as proposed in \citet{zilitinkevich2000} is a promising extension to MOST to account for the influence of independent ambient stratification in SBL. The fact that our value for $C=0.12$ is slightly smaller than the recommended $C=$0.2 in \citeauthor{zilitinkevich2000} could be due to the relatively small Reynolds number in our simulations.
It should thus be pointed out that
a separate investigation at higher Reynolds numbers with various combinations of Froude and stratification perturbation numbers is still required to verify the validity of Eq. (\ref{eq:phimextended}) with confidence.

\begin{figure}

\centering

    \includegraphics[width=\textwidth]{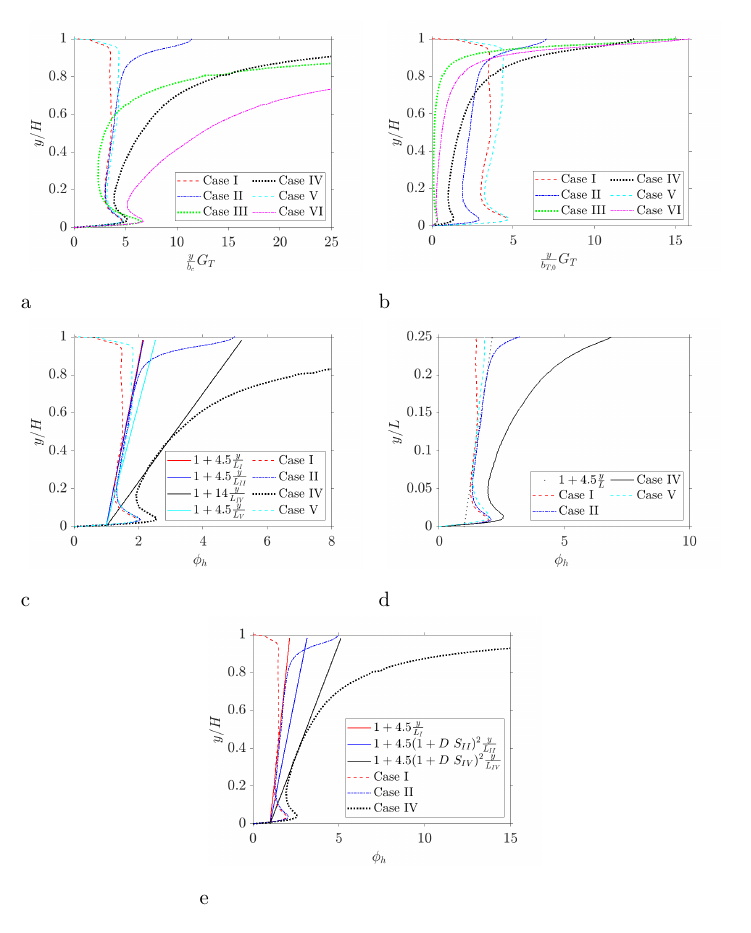}
\caption{ Plane-averaged total buoyancy gradient profiles along the channel height for differently stratified configurations  at quasi-stationary state: (a) Normalized by $b_c$; (b)  normalized by $b_{T,0}$;
  (c)  normalized by $b_*$  and compared to the original-MOST scaling with tuned coefficients;  (d) plotted in terms of stability parameter $y/L$ and (e) compared to the extended-MOST scaling \citep{zilitinkevich2000} with fitted constant $D=0.036$.   }
\label{fig:statmeanbmo}
\end{figure} 

\subsection{Analysis of mean buoyancy profiles}
In order to adequately account for coexistence of surface buoyancy flux with an independent ambient stratification, we introduce an alternative buoyancy scale which gives stronger weighting to the ambient stratification $N_a^2$. We note that this alternative scale is in addition to the previously established composite buoyancy $b_c=\beta N_c^2 \delta_{\nu}= b_*+N_a^2\beta/u_*$. This consideration is motivated by the fact that the independent ambient stratification is global, i.e. it operates at constant strength throughout the flow rather than being limited to a thin boundary layer. Our novel buoyancy scale is the
so-called reference \textit{ total buoyancy}  $b_{T,0}=b_*+b_N$, where $b_* = B_su_*^{-1} $ is the friction buoyancy based on the surface flux, whereas $b_N = N_a^2 h   $ is the \textit{reference bulk buoyancy} characterizing the ambient stratification, and $h$ is an equivalent boundary layer height which can be regarded as the minimal height above which the effect of surface wall becomes weakened enough such that turbulent effects are dominant. As can be seen from Fig. \ref{fig:statmeanub}b, for $y^{+}$ larger than $U_b/u_*$, which takes values between 14 and 18 for all stratified cases,  the log-law  region begins to be established. Therefore, $h=U_b/u_*\delta_{\nu}\approx 15 \delta_{\nu}$ would be a reasonable choice to evaluate $b_N = N_a^2 h$.    

Vertical profiles of the \textit{total buoyancy gradient} normalized by the composite buoyancy $b_c$ and total reference buoyancy $b_{T,0}$ for all stratified cases are shown in Figs. \ref{fig:statmeanbmo}a and \ref{fig:statmeanbmo}b. Here, based on the definitions given in Eq. \ref{eq:buoyancy} and the \ref{eq:bruntvaisala}, we define total buoyancy gradient $G_T$ as follows: 
\begin{align}
    G_T:= \frac{\partial<b>}{\partial y}+N_a^2.
\end{align}
We observe that while using $b_c$ as a buoyancy scale clusters the profiles of Cases \textit{I,II,V,VI}, normalization via $b_{T,0}$ clearly delineates the two Cases \textit{III,VI} containing solely ambient stratification from the remaining cases that have dominant surface fluxes. This effect is due to the stronger weighting of the ambient stratification contribution in the total reference buoyancy $b_{T,0}$ and suggests that it could be used to detect the presence of independent free atmosphere stratification from buoyancy gradient data. 

The vertical profile of the dimensionless total buoyancy gradient as defined in MOST via 
 \begin{equation}
 \phi_h=\frac{\kappa \ y}{b_*}G_T 
 \end{equation}
 is displayed in Fig. \ref{fig:statmeanbmo}c for all stratified cases except for cases \textit{ III,VI} which are stabilized purely by ambient stratification and for which the friction buoyancy $b_*$ computed from the surface flux is zero. 
 The empirical flux-profile relation for buoyancy gradient known from stable atmospheric boundary  layer studies \citep{businger1971,hogstrom1988} is: 
 \begin{equation}\label{eqnphih}
 \phi_h=\gamma_h + \alpha_h\frac{y}{L}
 \end{equation}
 This is the same empirical relation as for the dimensionless velocity gradient $\phi_m$, with different  coefficients
  $\alpha_h, \gamma_h$ that are chosen to optimally match empirical buoyancy gradient data. We set $\gamma_h=1.0$ like suggested in  \citet{hogstrom1988,shah2014},  and choose $\alpha_h$ to give the best fit to our DNS data. 
  We can see  from Fig. \ref{fig:statmeanbmo}c that the  relation implied in Eq. (\ref{eqnphih})  matches the linear near-surface portion of the simulated buoyancy gradient profile where $0.2~<~ y/H ~<~ 0.5$. In Fig. \ref{fig:statmeanbmo}d, we observe that the agreement between the linear empirical functions and the buoyancy profile breaks down at stability parameter $y/L ~>~ 0.22$ for $Case$ \textit{II} and at $y/L ~>~ 0.13$ for $Cases$ \textit{I, V}. Like for the velocity gradient profiles, the larger deviation of the buoyancy gradient from the linear profile  at higher regions could be attributed to the boundary condition imposed at the open channel top. 
  
  As discussed earlier for $\phi_m$, Fig. \ref{fig:statmeanbmo}c also indicates that the value of $\alpha_h$  depends sensitively on the strength of ambient stratification. DNS results for \textit{Case I}, which contains zero ambient stratification, gives \textcolor{black}{$\alpha_h=4.5$, which equals the corresponding values of $\alpha_m=4.5$ used to match the velocity gradient data.}  
   However, \textit{Case IV}, which has the strongest constant ambient stratification, requires a much larger $\alpha_h=14$, compared to $\alpha_m=7$ of case \textit{IV}.
   Thus,  these results indicate that existing empirical flux-profile relationship given by Eq. (\ref{eqnphih}) for mean buoyancy gradient needs to be modified to include the effect of prescribed ambient stratification that is not a product of the surface cooling.
   We refer to the extended MOST for buoyancy gradient given in \cite{zilitinkevich2000} which has the form 
\begin{equation}
\phi_h = 1.0 + \alpha_h(1  + D \cdot S)^2 \frac{y}{L}  \label{eq:phihextended}
\end{equation}
with dimensionless parameter  $S = N_a L u_{*}^{-1}$ and  empirical constant $D=1\pm 0.4$. We point out that since $\phi_h$ uses friction buoyancy $b_*=B_s u_{*}^{-1}$ to normalize the buoyancy gradient,   Eq. (\ref{eq:phihextended}) cannot be applied to cases without surface buoyancy flux which have $b_*=0$. This is in contrast to Eq. (\ref{eq:phimextended}) for velocity gradient which can be applied to  any combination of surface buoyancy flux and ambient stratification.
In order to match the lower portion of the buoyancy gradient profile of \textit{Case IV}, we determined  the value for $D$  to be around 0.036.  However, as shown in Fig. \ref{fig:statmeanbmo}e, this selected value for $D$ clearly over-predicts the normalized buoyancy gradient of \textit{Case II} with a weaker prescribed ambient stratification. These results demonstrate that, unlike its superior performance over the original MOST for velocity gradients, the extended similarity of \citeauthor{zilitinkevich2000} does not do a good job in predicting buoyancy gradients with a common set of constants. A separate investigation is needed to refine the formulation given in Eq. \ref{eq:phihextended}. 
 
\begin{figure}
\centering

\includegraphics[width=\textwidth]{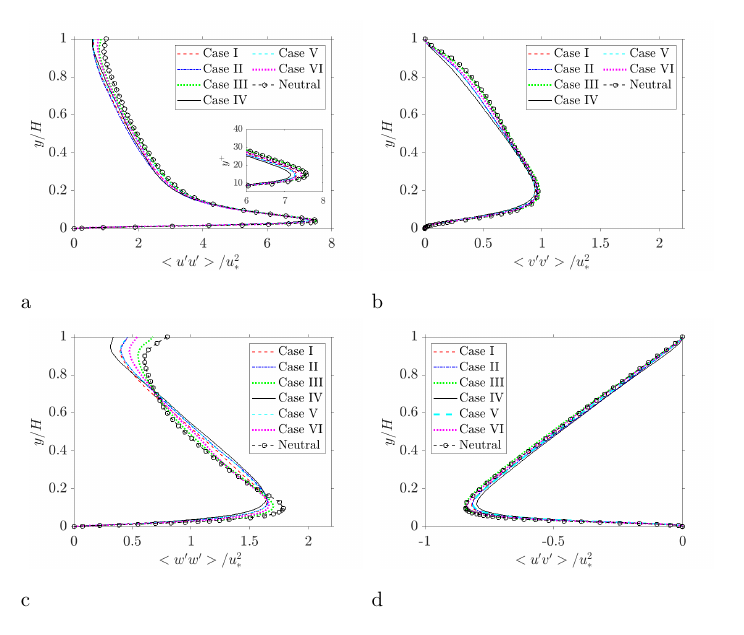}

\caption{ Vertical profiles of plane-averaged velocity variances and fluxes normalized by $u_{*}^2$ for different stratification configurations at quasi-stationary state:
(a) streamwise velocity component (inset: near-surface profiles as a function of $y^+$);
(b) vertical velocity component;  (c) transverse velocity component  and (d) turbulent vertical momentum flux.  }
\label{fig:statmeanvfluc}
\end{figure}
 
\subsection{Analysis of turbulence statistics}
 Profiles of plane-averaged normalized variances of all three velocity components as well as the main turbulent momentum flux across the channel height are shown in Fig. \ref{fig:statmeanvfluc}.
 We observe from Fig. \ref{fig:statmeanvfluc}a-\ref{fig:statmeanvfluc}b  that the vertical profiles for the streamwise and vertical velocity fluctuations are very similar  close to the surface  ($y/H ~<~ 0.2$) for all the differently stratified configurations, which could be explained by the dominance of mechanical turbulence over buoyancy there. However, as shown in Fig. \ref{fig:statmeanvfluc}c, the  normalized variance of the transverse (spanwise) velocity component $<w'w'>$ shows  far more noticeable variation near the surface than the normalized variances of other velocity components for the  stratified cases and the neutral flow: At $y/H \approx 0.1$, we observe that the magnitude of the $<w'w'>$ profile is clearly different for all cases, with the neutral flow achieving the largest value of around 1.75 and the most stable \textit{Case IV} reaching the smallest value around 1.55. 
  At higher locations within the channel core with $y/H ~>~ 0.2$, the stabilizing effect of stratification in suppressing the magnitude of vertical velocity fluctuations becomes more visible, as shown in Fig. \ref{fig:statmeanvfluc}b. We can also observe that within the open channel core where $0.2 ~<~ y/H ~<~ 0.6$, the normalized vertical velocity variances (Fig. \ref{fig:statmeanvfluc}b) of stratified flows are attenuated compared to neutral flow, whereas the normalized transverse velocity variances (Fig. \ref{fig:statmeanvfluc}c) are enhanced by stable stratification.
Therefore, we can state that turbulence anisotropy is enhanced with increasing stability within the channel core regions. 
 
\begin{figure}
\centering

    \includegraphics[width=\textwidth]{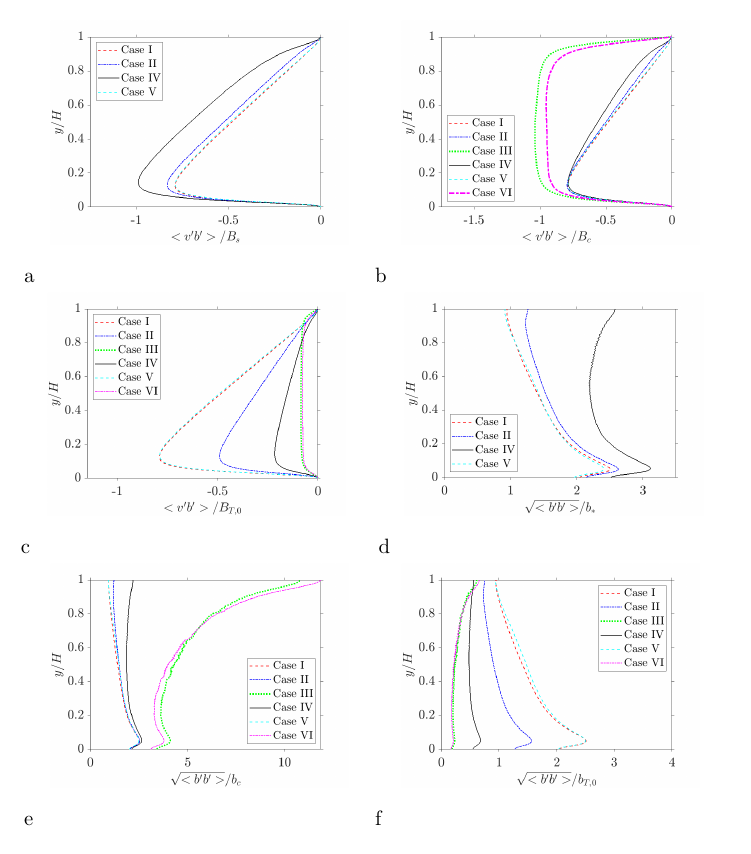}

\caption{ Plane-averaged turbulent buoyancy statistics  for various stratified configurations:  (a) vertical buoyancy flux normalized via the surface flux $B_s$; (b) vertical buoyancy flux normalized via the reference flux $B_c=\beta N_c^2 $;
(c)vertical buoyancy flux normalized via the reference flux $B_{T,0}=(b_*+b_N)u_* $;
(d) rms buoyancy fluctuation normalized via the friction buoyancy $b_*$; (e) rms buoyancy fluctuation normalized via reference buoyancy $b_c= N_c^2 \delta_{\nu} $;(f) rms buoyancy fluctuation normalized via reference buoyancy $ b_{T,0}=b_*+b_N$. 
\textit{Case III, VI} are excluded in  (a), (d) because they contain zero surface flux. }
\label{fig:statmeanbfluc}
\end{figure}
 
 The vertical profiles of normalized turbulent buoyancy fluxes and fluctuations for the stratified cases are displayed in Fig. \ref{fig:statmeanbfluc}. Under the present formulation with two independent stratification mechanisms, there is again the need to investigate which buoyancy scale is the most appropriate for normalizing turbulent buoyancy flux and fluctuation profiles.
 
 Three possible reference buoyancy fluxes come to mind for normalizing the turbulent flux profiles: (i) surface buoyancy flux $B_s$; (ii)  composite buoyancy flux $B_c =\beta N_c^2$ derived from composite stratification $N_c^2=N_a^2+G_w$ and (iii) total buoyancy flux $B_{T,0} = b_{T,0} u_*$ associated with the total buoyancy scale   $b_{T,0}=b_*+b_N$. The difference between $B_{c}$ and $B_{T,0}$ is that the latter scale \textcolor{black}{weighs} the contribution of ambient stratification more heavily.
 $B_s$ is not a suitable reference scale for cases \textit{III,VI} without surface flux. 
 
 Fig. \ref{fig:statmeanbfluc}a shows that the normalized flux profiles of configurations \textit{I, II,IV,V} with dominant surface fluxes would be separated by the strength of their respective ambient stratification $N_a$ components such that cases \textit{II,IV} with nonzero $N_a$ would have larger normalized vertical buoyancy flux magnitudes.  \textcolor{black}{Since a negative vertical buoyancy flux $<v'b'>$ is responsible for the destruction of turbulent kinetic energy in the TKE budget, the increase of its magnitude with growing overall stable stratification is to be expected.} This means that the presence of ambient stratification at a fixed surface flux would increase the vertical buoyancy flux magnitude, which is not accounted for in the purely surface-based scale $B_s$.  
  
 From Fig. \ref{fig:statmeanbfluc}b, we observe that choosing  the composite flux scale $B_c$  achieves the best collapse of the turbulent fluxes for cases \textit{I, II, IV, V} where surface buoyancy flux is significant ,  whereas the total buoyancy scale $B_{T,0}$ clearly separates those cases from each other as well as from the two cases \textit{III,VI} stabilized solely by ambient stratification (Fig. \ref{fig:statmeanbfluc}c). This indicates that the composite buoyancy $b_c$ may be a more suitable reference scale for buoyancy fluxes than the total reference buoyancy $b_{T,0}$.
 Figs. \ref{fig:statmeanbfluc}a and \ref{fig:statmeanbfluc}b both show that the buoyancy flux profiles  of cases where surface buoyancy flux is significant (\textit{I,II,IV,V}) have different shapes than those of the bulk-dominated stratified cases \textit{III,VI}: The former group achieves the maximal flux magnitude close to the surface and then decays rapidly further upwards, whereas the latter group, which has  zero  buoyancy flux at the bottom wall, retains its maximum turbulent flux  from $y/H~>~ 0.1$ until close to the  top of the channel at $y/H~>~0.9$. 
 As shown in Fig. \ref{fig:statmeanbfluc}b, the turbulent buoyancy flux profile of the bulk-dominated stratified cases \textit{III,VI}  normalized by $B_c$  attain their peak magnitudes, which are  larger and maintained at far higher locations until $y/H \approx 0.85$  than the other cases where the peak occurs at $y/H \approx 0.2$. This observation suggests that  turbulent thermal mixing processes in stable  flows can differ \textcolor{black}{noticeably} depending on the relative strength of the ambient stratification to surface flux as measured by $\Pi_s$; the presence of a prescribed ambient stratification appears to provoke a strong counter-gradient turbulent heat transfer to negate it.
 
 For the vertical profiles of root mean square (rms) buoyancy fluctuation $\sqrt{<b'b'>}$, we perform a similar analysis based on three possible buoyancy scales: (i) friction buoyancy scale $b_*=B_s /u_*$; (ii) composite buoyancy scale $b_{c}=N_c^2 \delta_{\nu} $;
 (iii) total buoyancy scale $b_{T,0}=b_*+b_N$. As can be seen by comparing Figs. \ref{fig:statmeanbfluc}d, \ref{fig:statmeanbfluc}e and \ref{fig:statmeanbfluc}f, the composite buoyancy $b_{c}$ is the most appropriate scale in collapsing the rms buoyancy fluctuation profiles for the dominant surface flux cases \textit{I,II,IV,V}   compared to the other two buoyancy scales, especially in the near-surface region. The total reference buoyancy $b_{T,0}$, however, clearly separates all  cases stabilized predominantly by surface flux from each other and from the pure ambient stratification cases \textit{III,VI}. This is in line with the observation for buoyancy fluxes as shown in Fig. \ref{fig:statmeanbfluc}b and \ref{fig:statmeanbfluc}c.
 When using  the friction buoyancy  to normalize the rms buoyancy fluctuation, we observe from Fig. \ref{fig:statmeanbfluc}d that the near-surface normalized variance values  for the most strongly stratified \textit{Case IV}  are around 50\% larger  compared to the more moderately stable \textit{Case I, II}.
 From Fig. \ref{fig:statmeanbfluc}e, it is clear that when normalized by  $b_{c}=N_c^2 \delta_{\nu}$, the bulk-dominant stratified cases \textit{III, VI} have significantly higher rms buoyancy fluctuations throughout the channel compared to the other cases, reaching values that are over 5 times larger at the peak. This agrees with the situation for turbulent buoyancy flux as shown in Fig. \ref{fig:statmeanbfluc}b where the buoyancy flux magnitude of cases \textit{ III,VI}  normalized by $B_c$ are also larger than the other cases. Thus, we conjecture that  large values of rms buoyancy fluctuations and buoyancy fluxes when normalized by the composite buoyancy scale $b_c$ may serve as an indicator to detect flow regions dominated by the influence of ambient stratification rather than surface cooling flux.
 Finally, Fig. \ref{fig:statmeanbfluc}f shows that when normalized by   $b_{T,0}$, the rms buoyancy fluctuation profiles of the bulk-dominated cases \textit{III,VI} are  close to each other, but there is greater separation for the other cases with dominant surface fluxes.
 
 The absence of a tight collapse for buoyancy related turbulence statistics near the surface and within the channel core  is well contrasted by  the close  agreement of normalized velocity statistics profiles for all the  cases, as shown in Figs. \ref{fig:statmeanvfluc}a-\ref{fig:statmeanvfluc}d. 
 This suggests that with sufficient strength of ambient stratification, as quantified by a $\Pi_s$ value not much larger than unity, buoyancy effects on turbulence become significantly altered throughout the channel. Its analysis may benefit from an additional reference buoyancy scale and an extended similarity theory to characterize its full effect. 

Despite the different options  to normalize the turbulent buoyancy flux or rms buoyancy fluctuation, both quantities, independent of the choice of normalizing scale, can be used to quantify the strength of the stratification in the bulk core. To this end, with the exception of idealized cases {III} and {VI} that lack any surface cooling, the progression from the least stable \textit{Case I} toward the most stable \textit{Case IV} is evident.

 \subsection{Further analysis of stability effects on turbulence}
\begin{figure}

    \includegraphics[width=\textwidth]{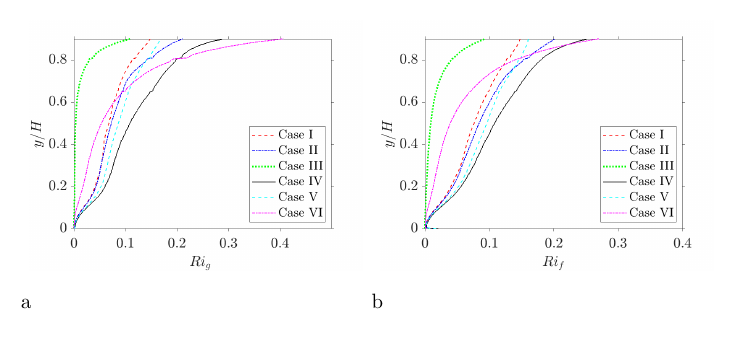}
  
\caption{Quantification of flow stability for different stratification configurations: Vertical profiles  of  (a) gradient Richardson number $Ri_g$ and  (b) flux Richardson number $Ri_f$.}
\label{fig:statmeanRigtke}
\end{figure}

 To obtain a further understanding of flow stability and its quantification, we present in Figs. \ref{fig:statmeanRigtke}a \& \ref{fig:statmeanRigtke}b, the vertical profiles of the gradient Richardson number ($Ri_g$) and flux Richardson number ($Ri_f$), respectively. These dimensionless parameters are defined via the mean flow profiles as
 \begin{equation}\label{eqnRig}
     \text{Ri}_g=\frac{\partial <b>/\partial y +N_a^2}{(\partial <u>/\partial y)^2}=\frac{G_T}{{(\partial <u>/\partial y)^2}},  \qquad Ri_f = \frac{<v'b'>}{\frac{\partial <u>}{\partial y} <u'v'>}. 
 \end{equation}

We observe that in all cases, $Ri_g$ and $Ri_f$ both stay less than the commonly acknowledged critical value of 0.2 for most of the lower channel up to $y/H=0.8$, which should justify the application of local similarity theory according to \citet{grachev2007}. However, as we have already seen in Figs. \ref{fig:statmeanmo}a  and \ref{fig:statmeanbmo}c,  for channel heights $y/H ~>~ 0.1$ onward, the mean gradients of cases with significant ambient stratification, i.e. \textit{Cases IV,  VI}, clearly depart from the MOST, which otherwise accurately agreed with the profiles of pure surface flux flows, i.e. \textit{Cases I,V}.     
 
\begin{figure}
\centering
\includegraphics[width=\textwidth]{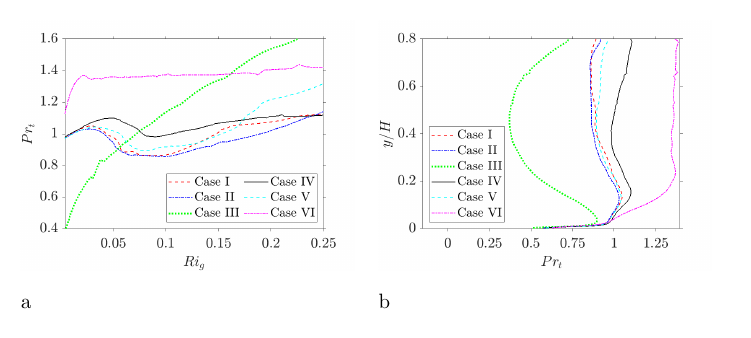}
  \caption{   Turbulent Prandtl number $Pr_t$ as a function of (a) gradient Richardson number $Ri_g$  and (b) dimensionelss height $y/H$.}
\label{fig:NhcomparePr}
\end{figure}

 The ratio between the gradient and flux Richardson numbers, which also equals the ratio between turbulent viscosity and diffusivity, is known as the turbulent Prandtl number $Pr_t=Ri_g / Ri_f$ and plotted in Fig. \ref{fig:NhcomparePr}a as a function of the gradient Richardson number $Ri_g$. 
 The dependence of $Pr_t$ on $Ri_g$ has been documented in many other numerical and experimental studies \citep{armenio2002channel,grachev2007}. It is generally accepted that for stratified flows, $Pr_t$ is approximately unity in regions of low stability characterized by $Ri_g<0.2$. We can observe from Fig. \ref{fig:NhcomparePr}a that this holds true for all configurations except for the two cases stabilized purely by ambient stratification: \textit{Case III} attains  values as low as \textcolor{black}{$Pr_t \approx 0.4$} for $Ri_g\approx 0$, whereas \textit{Case VI} achieves  $Pr_t > 1.15$ throughout $Ri_g< 0.2$.  Since \textit{Case III}, which only contains a weak ambient stratification, is known to be the weakest stable case at almost neutral conditions, its turbulent Prandtl number can deviate substantially from unity according to \citet{li2015}. On the other hand, the turbulent Prandtl of the cases \textit{I,II,IV,V} with dominant surface fluxes ($\Pi_s>1$) all satisfy $0.85 ~<~ Pr_t ~<~ 1.10$ throughout the shown region.  However, among those cases, clear differences are visible in the  stability region where $0.04 ~<~ Ri_g ~<~ 0.15$: For the cases \textit{I, II ,V} with very weak ambient stratification ($\Pi_s \gg 1$) the turbulent Prandtl number stays below 1.05, whereas the value of $Pr_t$ for \textit{Case IV}, where ambient stratification is more substantial with $\Pi_s=4$,   is clearly larger.
  We can also see for the cases \textit{I,II,V} that $Pr_t$ decreases with growing gradient Richardson number for $0.05~<~ Ri_g ~<~ 0.08$ before increasing towards 1 at larger $Ri_g$. For \textit{Case IV}, the exact opposite behavior of $Pr_t$ as a function of $Ri_g$ can be observed. This trend is not observed in  \citet{armenio2002channel} which studied stratified flows with surface cooling only, hence could be attributed as another effect of ambient stratification on turbulent buoyancy quantities.
 
 The vertical profile of the turbulent Prandtl number in the lower half of the channel is shown in Fig. \ref{fig:NhcomparePr}b, which indicates that  for all configurations except for \textit{Case III},  $Pr_t\approx 1$  is attained at $y/H \approx 0.05$. For \textit{Case III}, however, $Pr_t<0.8$ in that near-surface region, significantly lower than the other cases. This agrees with the deviation shown in the low $Ri_g$ portion of Fig. \ref{fig:NhcomparePr}a and is a manifestation of the anomalous character of the very weakly stratified flow, as described earlier and pointed out by \cite{li2015}. It is however of interest to observe that the three cases \textit{I,II,V} predominantly stabilized by surface flux with $\Pi_s\gg 1$ have very similar Prandtl number profiles, whereas the other two cases \textit{III,VI} stabilized solely via ambient stratification contribution deviate from each other markedly. 

\begin{figure}
\centering

    \includegraphics[width=\textwidth]{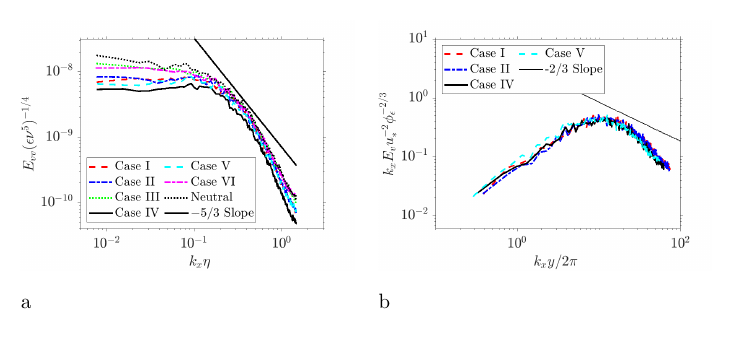}

\caption{  Streamwise spectra of the vertical velocity fluctuations at quasi-stationary state:  (a) at open-channel height $y/H=0.5$; (b) pre-multiplied  spectra  for the weakly stable regime $y/L=0.1$. Cases \textit{III, VI} have been excluded  in (b) because they contain zero surface flux and have ill-defined Obukhov length $L$.    }
\label{fig:spectrvcompare}

\end{figure}

Normalized streamwise spectra of vertical velocity for different flow configurations are shown in Fig.  \ref{fig:spectrvcompare}.
In Fig. \ref{fig:spectrvcompare}a,
the  spectra of \textcolor{black}{wall-normal} velocity  for all stratified configurations and the neutral flow are plotted against the  wavenumber, all  normalized by the Kolmogorov scale, at channel height $y/H=0.5$.  We observe from  Fig. \ref{fig:spectrvcompare}a that the energy content at all scales decreases with increasing degrees of
stable stratification.  At the largest scales, the energy stays nearly constant , whereas it decays approximately with  -8/3  power at the smallest scales (\textcolor{black}{dissipation range}) with growing wave numbers. The near-constancy of energy at the lowest wave numbers can be attributed to the randomized distribution of energy among all scales for wave numbers below a threshold \textcolor{black}{since eddies of sizes comparable to the wall distance may interact with the solid boundary}, as explained by \citet{li2015,katul2014,grachev2013}.
All spectra follow the -5/3 decay within the intermediate narrow inertial range, hence suggesting that continuous, fully developed turbulence has been achieved in all cases studied herein. 

In the study of atmospheric boundary layers, it has been deemed beneficial to normalize the  flow spectra  with quantities that appear in MOST. As demonstrated in \citet{kaimal1994}, the so-called pre-multiplied spectra  at different stability parameters $y/L$ all collapse in the inertial subrange, where the decay rate follows a $-2/3$ slope, but bifurcate in the smaller wave numbers. 
Fig. \ref{fig:spectrvcompare}b shows the pre-multiplied  spectra at the  stability parameter $y/L=0.1$ for all stratified flow cases except for \textit{Case III,VI}, which have very weak surface fluxes. The spectra of \textit{Case I,II}, which differ in their setup only by the small strength of ambient stratification imposed in \textit{Case II}, overlap each other nearly perfectly. This demonstrates that the small magnitude of ambient stratification in \textit{Case II} has negligible effect on the vertical velocity  fluctuations, in agreement with the  velocity variance profile displayed in Fig. \ref{fig:statmeanvfluc}b.
We observe that the spectra at the same stability \textcolor{black}{collapse onto each other to a good degree, but not as perfectly}  as predicted in \citet{kaimal1994}. The small deviations in spectra may be attributed to the relatively small Reynolds number $Re=360$ and narrow inertial ranges of the flows studied here.

\section{Conclusions}
We have investigated the stabilizing effects of independent and simultaneous action of two stratification mechanisms, namely a constant negative surface buoyancy flux and a prescribed ambient stratification, on an initially neutral turbulent open-channel flow through direct numerical simulations. Our major findings are as follows:
\begin{enumerate}

   \item A flow configuration with prescribed ambient stratification as well as surface cooling is controlled by four independent dimensionless parameters: $Re_b, Fr_b, \Pi_s, Pr$.  The stratification perturbation number $\Pi_s$ relates the strength of surface cooling to the prescribed ambient stratification. $\Pi_s$ enters the dimensionless parameter space only when both stratification mechanisms are active. As a rule of thumb, the overall stability of a flow with separate contributions from surface cooling and ambient (background) stratification can be determined through the composite Froude number $Fr_c$, which is a function of $Fr_b$ and $\Pi_s$. And at constant $Fr_c$, the larger the inverse $\Pi_s$, the stronger the stability. In the linearized governing equations (see section \ref{sec:linstable}), both types of stratification mechanisms are interchangeable, thus $Fr_b$ and $\Pi_s$ always appear together in a unified form as $Fr_c$ in the linear stability equations (\ref{eqnslopelincont})-(\ref{eqnslopelinlast}). However, $\Pi_s$ appears as an independent parameter in the dimensionless form of the Navier-Stokes equations, and
   influences the flow field through nonlinear interactions as shown via our DNS results.

    \item Despite the common agreement that Monin-Obukhov similarity theory (MOST) holds for weakly stratified flows with gradient or flux Richardson numbers below 0.2, we show through DNS that this expectation needs to be modified when the source of flow stratification is a mechanism other than surface cooling, such as an independent ambient stratification. Flux-profile relationships exhibit a clear dependence on the strength of independent ambient stratification, which is captured rather well in the extended MOST proposed by \cite{zilitinkevich2000}, requiring an additional empirical constant $C$ to model the effect of ambient stratification. This extended MOST relationship matches the near-surface portion of the  normalized velocity gradient profiles in our simulations with $C=0.12$, whereas the suggested value for $C$ in \citet{zilitinkevich2000} is 0.2. However, the velocity gradients at higher elevations in our simulations deviate from the  profile specified by the extended MOST. This deviation is also dependent on the type of the thermal boundary condition imposed at the open channel top boundary (see Appendix \ref{sec:effectbc}). We also observe a less accurate agreement between our simulated buoyancy gradients and the extended similarity theory. These observations suggest that additional investigations still need to be done to arrive at a more accurate extension to MOST that is capable of reliably predicting the gradient and fluxes in flows stratified with multiple independent mechanisms.
    
    \item Turbulent statistics of buoyancy in stratified flows where independent ambient stratification is dominant behave differently than flows stratified by surface cooling only, even in the weakly stable regime: The normalized buoyancy fluxes and rms fluctuations at higher distances from the surface are  stronger, and 
    turbulent Prandtl number near the surface clearly deviates from the expected value of 1.0. This suggests that those quantities could be used to gauge for the presence of an independent ambient stratification in the atmospheric boundary layer.
    However, the signature of ambient stratification on turbulent statistics of the velocity field is far less pronounced and, therefore,  might not be as useful to infer the source of stratification.  
    
    \item Different scales are possible to normalize the profiles of turbulent buoyancy flux or the rms buoyancy fluctuation: The friction buoyancy $b_*$ is a suitable scale near the surface but fails to account for the effect of independent ambient stratification in the open-channel core, whereas composite buoyancy scale ($b_{c}=N_c^2 \delta_{\nu} $) incorporates the overall stratification effect of both mechanisms, but it gives insufficient weight to the ambient stratification. Motivated by the observation that ambient stratification has a far stronger stabilizing effect than a stable surface flux of the same magnitude, another novel total buoyancy scale $ b_{T,0}$ is introduced here, which gives a stronger weight to the independent ambient stratification. This scale seems to be adequate to normalize and bring the buoyancy profiles of all configurations within a reasonable range. However,  in contrast to applying the composite buoyancy scale $b_c$, profiles of turbulent buoyancy statistics where multiple stratification mechanisms are present do not tightly collapse to a common curve when normalized by $b_{T,0}$, pointing to a need for future investigations. 

\end{enumerate}

Finally, a major implication of our investigation pertaining to laboratory- and field-scale experiments on stratified flows is the following: \textit{How a flow gets stratified matters!} We argue that simply quantifying the final stratification through the Brunt-Vais\"al\"a frequency is not sufficient. It is critical to elucidate the mechanisms underlying the resulting bulk stratification. This task is straightforward in the laboratory as the method to stratify the flow is known ahead of the measurements. However, the issue is expected to be more involved and uncertain when it comes to inferring the mechanisms behind an observed stratification in the field. 
Major departure from MOST under weakly stable conditions or response of the turbulence statistics of buoyancy to different scaling parameters can be used as a hint that stratification mechanisms other than surface cooling might be at play. 

\begin{acknowledgments}
Research was sponsored by the National Science Foundation under Award Number \textbf{1936445} and in part by the University of Pittsburgh Center for Research Computing through the resources provided. This work used the Extreme Science and Engineering Discovery Environment (XSEDE), which is supported by National Science Foundation grant number ACI-1548562, and the Expanse cluster at the San Diego Supercomputer Center through allocation TG-ATM200017.
\end{acknowledgments}

\appendix

\section{\textcolor{black}{Effect of  thermal boundary condition at the open channel top boundary}}\label{sec:effectbc}
\textcolor{black}{The thermal  condition to be applied at the top boundary in numerical studies of ABL has been subject to ambiguity since no authoritative guideline pertaining it has been established yet. Two of the most commonly used approaches are Dirichlet and Neumann boundary conditions: The former fixes a constant temperature or density value at the top whereas the latter imposes an adiabatic top  boundary. In the  simulations of stratified open channel flows by \cite{nieuwstadt2005,flores2011} and of stratified Ekman layers by \cite{deusebio2014}, temperature has been fixed at the top boundary, whereas the Neumann condition for temperature or density has been applied in the numerical studies by \cite{taylor2005,Gohari2018,shah2014}. A comparative study for these two types of boundary conditions in open channel flows has been conducted in \cite{atoufi2020characteristics}. From our current results, we also aim to examine the effect of different thermal conditions at the  channel top boundary on turbulent statistics and flow profiles for  cases with multiple stratificaton mechanisms.  }

\begin{figure}
\centering

    \includegraphics[width=\textwidth]{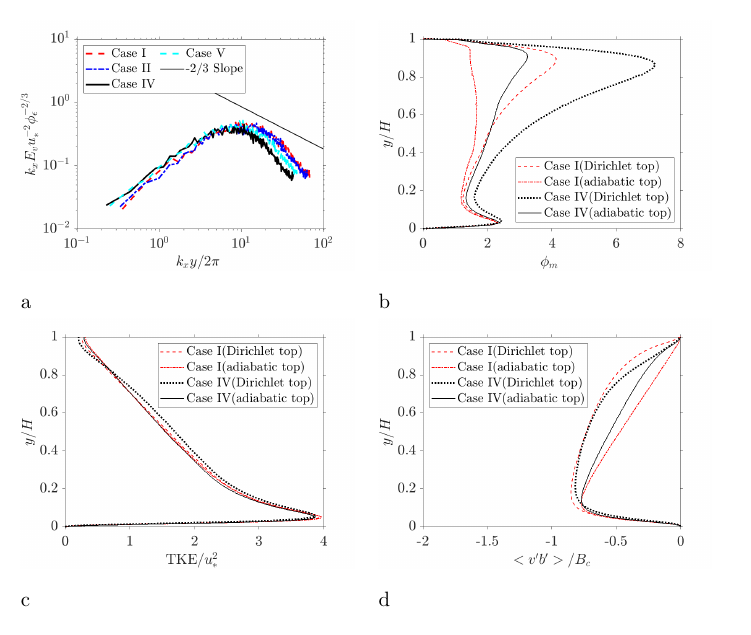}

\caption{  Effect of the choice of thermal boundary condition at the open channel top boundary: (a) pre-multiplied  spectra  for the weakly stable regime at stability parameter $y/L=0.094$ for Dirichlet boundary condition of $b=0$;(b) comparison of plane-averaged normalized shear profiles (c)  TKE profiles and (d) buoyancy flux profiles when applying adiabatic or Dirichlet ($b=0$) boundary condition.}
\label{fig:bch0compare}

\end{figure}

\textcolor{black}{
In Fig. \ref{fig:bch0compare}a, the pre-multiplied streamwise spectra  for the wall-normal velocity fluctuations of cases \textit{I,II,IV,V} with Dirichlet thermal boundary condition at the top are for stability value $y/L=0.094$. Compared to the corresponding Fig. \ref{fig:spectrvcompare}b for the same cases but with adiabatic top boundary, we can see that their spectra do not collapse when evaluated at the same weakly stable regime, which points to a distorting influence of the Dirichlet thermal boundary condition on the velocity fluctuations. This is also supported when comparing normalized shear profiles of cases \textit{I,IV} with different top boundary conditions as displayed in Fig. \ref{fig:bch0compare}b, showing that configurations with Dirichlet top thermal boundary condition experience larger shear from wall distance $y/H>0.2$ onwards. A plot of the plane-averaged TKE profiles (Fig. \ref{fig:bch0compare}c) also shows that fixing the buoyancy value at the top channel boundary increases the TKE in the outer layer compared to the adiabatic top boundary, especially in the presence of ambient stratification as in case \textit{IV}. This qualitative behavior of the TKE profile depending on the channel top boundary agrees with the findings in \cite{atoufi2020characteristics}.
 The turbulent buoyancy statistics as shown in Fig. \ref{fig:bch0compare}d also demonstrate that the turbulent buoyancy fluxes  are far larger from the outer layer upwards when  a Dirichlet thermal boundary condition is applied at the top, which can be attributed to heat entrainment due to the fixed  temperature at the channel top \cite{atoufi2020characteristics}.
}

\section{Linear  Flow Characterization  }\label{sec:linstable}
We carry out a linear stability analysis to better understand the role of the dimensionless parameter space on the dynamical stability of the flow configuration. We note that nonlinear and non-modal effects can be important in dynamical stability of fluid flow systems and they cannot be studied with a linear stability analysis. 

We linearize the Navier-Stokes equations (Eqs. \ref{eqnslopemom}-\ref{eqnslopebuoy} and \ref{eqnslopecont}) around a parabolic velocity profile and a linear  buoyancy profile with gradient $G_w$ for a laminar open channel flow  subject to constant ambient stratification $N_a$. As part of the method of normal modes we assume that disturbances to the base flow are waves of the form $\mathbf{q}(x,y,z,t)=\mathbf{\hat{q}}(y) \exp{\{ i(k_x x+k_z z) + \omega t \}}$, where $i$ is the imaginary unit. $k_x, k_z$ are real wavenumbers in the $x$ (streamwise) and $z$ (spanwise) directions, respectively, whereas $\omega$ is a complex frequency. Note that, gravity acts in the $y$ direction in our formulation. The resulting system of linear stability equations normalized by the flow scales given in Eq. (\ref{eq:scales}) can be written as follows:
\begin{align}
ik_x\hat{u}+ik_z \hat{w}+ \frac{\partial \hat{v}}{\partial y} &= 0,\label{eqnslopelincont}\\
\omega \hat{u}+iu_nk_x\hat{u}+u_n'\hat{v}
&= -ik_x\hat{p}+\frac{ 1}{\text{Re}_b}\left(-(k_x^2+k_z^2)\hat{u}+\frac{\partial^2\hat{u}}{\partial z^2} \right),\\
\omega \hat{v}+iu_nk_x\hat{v}
&= -\frac{\partial\hat{p}}{\partial y}+\frac{ 1}{\text{Re}_b}\left(-(k_x^2+k_z^2)\hat{v}+\frac{\partial^2\hat{v}}{\partial y^2}\right) + \frac{1}{\text{Fr}_c^2}\hat{b}   , \\
\omega \hat{w}+iu_nk_x\hat{w}
&= -ik_y\hat{p}+\frac{ 1}{\text{Re}_b}\left(-(k_x^2+k_z^2)\hat{w}+\frac{\partial^2\hat{w}}{\partial y^2} \right),\\
\omega \hat{b}+iu_nk_x\hat{b}+\hat{v}
&= \frac{ 1}{\text{Re}_b\text{ Pr}}\left(-(k_x^2+k_z^2)\hat{b}+\frac{\partial^2\hat{b}}{\partial y^2}  \right) ,\label{eqnslopelinlast}
\end{align}
where $\hat{u},\hat{v},\hat{w},\hat{p},\hat{b}$ are  flow disturbances varying in the vertical $y-$ direction. The normalized  base flow solution and its derivative in the vertical direction are represented as $u_n,b$ and $u_n'$, respectively.

It is noteworthy that, in contrast to the dimensionless form of the Navier-Stokes equations (Eqs.  \ref{eqnslopemom}-\ref{eqnslopebuoy} and \ref{eqnslopecont}), the dimensionless linear stability equations (Eqs. \ref{eqnslopelincont}-\ref{eqnslopelinlast}) do not contain the stratification perturbation number $\Pi_s$ explicitly, even though the same flow scales have been chosen to normalize both sets of equations. However, we observe that only the composite Froude number $Fr_c$, thus the composite stratification $N_c^2$, is of significance in the linearized problem. If we recall the definition of $N_c$ from Eq. (\ref{eq:Nc}) we can deduce that configurations with the same $N_c$ but different values for $N_a$ and $G_w$ should result in the same linear behavior, meaning that the strength of the surface heat flow relative to the ambient stratification, as quantified by $\Pi_s$, does not have a unique separate role in the linear flow problem. However, as  our direct numerical simulation results  have shown, the situation is quite different in the nonlinear flow problem where the strengths of each stratification mechanism do matter separately. This is further evidenced by the appearance of both $Fr_c$ and $\Pi_s$ separately in Eqs. (\ref{eqnslopemom}) and (\ref{eqnslopebuoy}).

\begin{figure}

    \includegraphics[width=\textwidth]{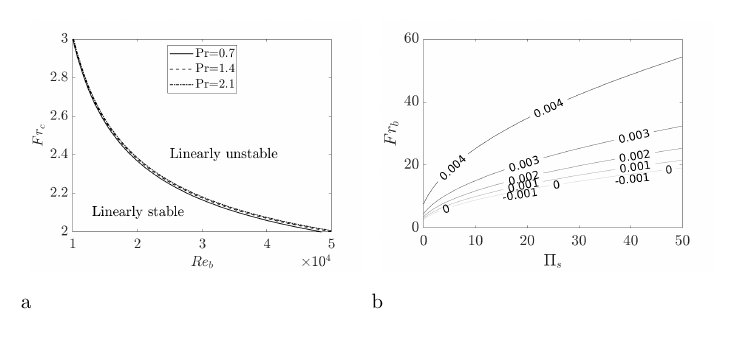}

\caption{Linear stability behavior for the stably stratified laminar channel flow containing both surface buoyancy flux and ambient stratification:
  (a) Critical composite Froude number $F_c$ of linear stability as function of  bulk Reynolds number $Re_b$   for  different Prandtl numbers and (b) maximal growth rate contour as function of bulk Froude number $F_b$ and stratification perturbation number $\Pi_s$ at  bulk Reynolds number $Re_b=20,000$. \textit{These growth rate contours are also contours of constant composite Froude number $Fr_c$ which alone determines flow stability in the linear case.}}
\label{fig:stabReFrN}
\end{figure}

We solve the linear stability equations for the stratified channel flow configuration with imposed surface buoyancy flux (Eqs. \ref{eqnslopelincont}-\ref{eqnslopelinlast}), using pseudo-spectral discretization as described in \cite{schmid2012book}. 
To illustrate how different dimensional numbers impact the degree of stability for a stratifed flow,
we display the  neutral stability curves in terms of the composite Froude number $Fr_c$ and bulk Reynolds number $Re_b$  in Fig. \ref{fig:stabReFrN}a.
 The stability plot in Fig. \ref{fig:stabReFrN}a  indicates that the Prandtl number, over the range that it was investigated, has little influence on the critical composite Froude number and  exhibits trends similar to the trends reported in \citet{gage1968}, who stipulated constant temperature differences instead of surface buoyancy flux for stratification of the laminar channel flow.
From Fig. \ref{fig:stabReFrN}a, it is evident that the critical composite Froude number declines with growing Reynolds number. Fig. \ref{fig:stabReFrN}b shows the maximal growth contours for a given bulk Froude number $Fr_b$ and stratification perturbation number $\Pi_s$, which are also contours of constant composite Froude number $Fr_c$ . We observe that the growth rate increases with growing $Fr_b$ at a fixed $\Pi_s$ due to the decrease in stable prescribed ambient stratification; however, at a constant $Fr_b$, increasing $\Pi_s$ implies a higher stable surface flux, which leads to a diminishing of the growth rate.

The  plots  of Fig. \ref{fig:stabReFrN} demonstrate that under laminar conditions, the stability of stratified flows can  be fully characterized using either the composite Froude number $Fr_c$ or a combination of the stratification perturbation $\Pi_s$ and bulk Froude number $Fr_b$: Everything else being equal, the smaller value of either Froude number indicates a more stable flow, and increasing $\Pi_s$ implies  less stable conditions.
However, it should be cautioned that Figs. \ref{fig:stabReFrN}a and  \ref{fig:stabReFrN}b only serve a qualitative purpose to understand parameter dependence of dynamic stability. Linear stability theory has been \textcolor{black}{shown  not to provide} quantitative predictions about laminar-turbulent transitions in channel flows \citep{schmid2012book}. 

\section{Impact of flow forcing method on post-stratification turbulence}\label{sec:bulkdp}

To contrast the evolution of neutral turbulence under stabilizing buoyancy flux at constant pressure gradient versus constant mass flow rate, we carried out two DNS, both starting from the same initial neutral open channel flow with $Re_{\tau 0}=180$, and imposed on them the same surface cooling flux that gives a Froude number of $Fr_c = 2.24$. No prescribed ambient stratification has been applied in either case.  The first simulation drives the flow with the same streamwise pressure gradient of the initial flow, whereas the second simulation uses the same initial flow rate as forcing. 
To facilitate comparisons, we will use the initial friction velocity $u_{*0}$ to define the dimensionless time $t_*=t/u_{*0}$ for both cases.
\begin{figure}

    \includegraphics[width=\textwidth]{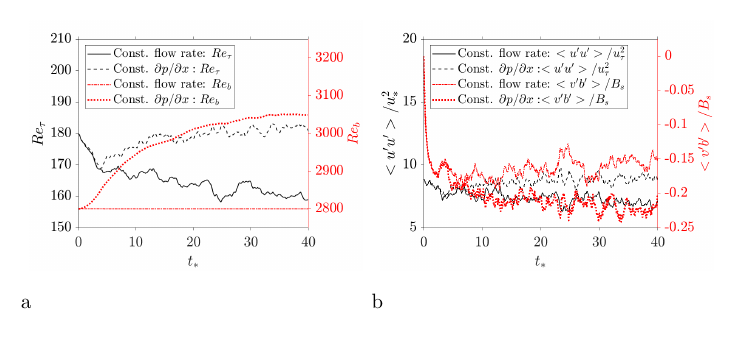}

\caption{Comparison between  flow quantities evolution from initial neutral flow at  $Re_{\tau}=180, Pr=1.0$ after onset of stable surface buoyancy flux at constant pressure gradient and constant flow rate: (a) Friction Reynolds number and bulk Reynolds number; (b) Streamwise velocity variance and vertical turbulent buoyancy flux at $y/H=0.02$. }
\label{fig:pmcomparetime}
\end{figure}
\paragraph{Evolution of turbulent quantities} 
For both simulations, the development of friction Reynolds number $Re_{\tau}=u_{*}H/\nu$ and bulk Reynolds number $Re_b=U_b H/\nu$  over a time interval of $t_*=40$ are shown in Fig. \ref{fig:pmcomparetime}a. In agreement with expectations, we observe that at constant flow rate, $Re_{\tau}$ shows a clear decreasing trend until reaching a statistical plateau $Re_{\tau}=159$, which is approximately 88\% of its initial value of $Re_{\tau 0}=180$  at around $t_*=35$. By contrast, in the simulation driven by constant pressure gradient, the friction Reynolds number returns to the initial value $Re_{\tau 0}$ at $t_*=15$ after an initial decrease. It  noteworthy that despite the adoption of different flow forcing techniques,  both cases exhibit an almost identical decrease in $Re_{\tau}$ up until $t_*=4$. 
Bulk Reynolds number shows a monotonous increase for the pressure-driven case until $t_*=35$ when it reaches a stationary plateau, which is roughly 9\% higher than its initial value of $Re_{b,i}=2800$. 
The increase  in $Re_b$ could therefore be seen as offsetting the stabilizing effect of surface cooling and hence responsible for the resurgence  of $Re_{\tau}$ in the pressure-driven case.

The evolution of streamwise velocity variance and turbulent buoyancy flux near the surface at $y/H=0.02$ is shown in Fig. \ref{fig:pmcomparetime}b. Following the  development of friction Reynolds number discussed earlier, we observe that turbulent quantities for both cases display a similar downward trend until $t_*=5$ independent of  the type of the flow forcing. However, in the pressure-driven case, the velocity fluctuation begins to increase after this initial decay until recovering its initial value at $t_*=15$, mimicking the behavior of its friction Reynolds number. A relatively weaker recovery trend can be observed for its turbulent buoyancy flux, which, however, stays negative throughout. In the flow-rate driven case, however, turbulent quantities continue their initial downward trend beyond $t_*=5$ until settling down to statistical stationarity, with weaker velocity fluctuations but stronger negative turbulent buoyancy flux.

\begin{figure}
\centering

    \includegraphics[width=\textwidth]{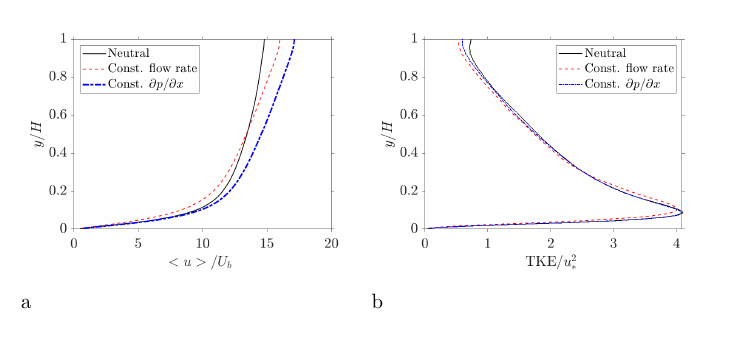}
  
\caption{Comparison between plane-averaged  flow profiles along the vertical direction at $t_*=40$ after onset of stable surface buoyancy flux under the action of different flow forcing techniques, along with the initial neutral velocity profile : (a) Streamwise velocity normalized by bulk quantities ; (b) Turbulent kinetic energy normalized by friction velocity.}
\label{fig:pmcompareprofile}
\end{figure}

\paragraph{Stationary flow profiles} 
From Fig. \ref{fig:pmcomparetime}, we observe that both cases of stratified flows have evolved toward a new statistically stationary state after $t_*=40$. The plane-averaged stream-wise velocity profile for both cases are shown in Fig. \ref{fig:pmcompareprofile}a along with the initial neutral flow profile. The effects of each forcing type are clearly visible: For the stratified flow at constant flow rate,  deceleration
occurs near the surface due to the diminished wall shear caused by stable buoyancy flux. However, the amount of velocity reduction gradually decreases with growing height and reverses into acceleration for $y/H>0.5$, caused by the diminishing momentum transfer due to stable stratification. On the other hand, the flow driven by constant pressure gradient  accelerated throughout the channel by a larger magnitude due to the increased mass flow rate. 
The  magnitude of flow acceleration as a result of  flow forcing type may have implications in studying turbulence cessation in stratified flows. For instance, \citet{vanwiel2012minimum} have suggested that the magnitude of maximum velocity in stably stratified atmospheric flows is of significance for turbulence collapse.

Despite these clear differences in mean velocity profiles, Fig. \ref{fig:pmcompareprofile}b shows that the normalized turbulent kinetic energy (TKE) profiles for all three cases agree very well with each other, especially near the surface. Smaller differences only begin to show up at $y/H>0.5$ where reductions in TKE due to stable stratification relative to the neutral case become visible. This trend suggests that the stratified cases studied thus far are only weakly stable and far from the point of flow laminarization, which is also supported by plots shown in Fig. \ref{fig:pmcomparetime} where we can observe that the  friction Reynolds number and streamwise velocity variance never fall below 80\% of the initial value.  

\end{document}